\documentclass[letterpaper,twocolumn,10pt]{article}
\usepackage{usenix-2020-09}

\usepackage{tikz}
\usepackage{amsmath}
\usepackage[usestackEOL]{stackengine}
\usepackage{floatrow}
\usepackage{verbatim}
\usepackage{amsmath}
\usepackage{algorithm}
\usepackage{algorithmic}
\usepackage{caption}
\usepackage{graphicx}
\usepackage{lipsum}

\usepackage{filecontents}

\usepackage{glossaries}
\usepackage{mathtools}

\begin{document}


\title{EaCO: Resource Sharing Dynamics and Its Impact on Energy Efficiency for DNN Training}


\author{
{\rm Kawsar Haghshenas}\\
University of Groningen\\
k.haghshenas@rug.nl
\and
{\rm Mona Hashemi}\\
University of Tehran\\
hashemi.mona@ut.ac.ir
} 

\maketitle

\begin{abstract}
Deep Learning Training (DLT) is a growing workload in shared GPU/CPU clusters due to its high computational cost and increasing number of jobs. This contributes to significant energy consumption in GPU clusters, further exacerbated by GPU under-utilization, as shown in production cluster logs. Addressing this challenge requires workload scheduling and resource allocation policies for efficient GPU sharing to improve resource and energy efficiency while maintaining performance. However, previous works primarily optimize for performance, often overlooking or even sacrificing energy efficiency.

In this paper, we present EaCO, the first energy-aware scheduling algorithm designed specifically for DLT workloads in GPU clusters. EaCO leverages hardware-supported context switching to enable GPU sharing across multiple DLT jobs, improving resource and energy utilization. GPU sharing can increase Job Completion Time (JCT) and may lead to contention if not employed carefully. To address this, EaCO integrates experiment and historical-based predictions as well as early-stage observations, ensuring performance expectations are met while optimizing energy efficiency.

We begin by experimentally exploring the dynamics of co-locating DLTs, investigating its impact on energy and resource utilization. Our results show that co-location improves energy efficiency by up to 44\% for individual jobs, and increases average GPU utilization to as high as 97\%. Additionally, evaluations on large-scale clusters using production traces demonstrate that EaCO reduces total energy by up to 39\% compared to existing algorithms, which comes with a minimal increase in job runtime—less than 3.2\% in our simulations.

\end{abstract}

\section{Introduction}
Deep Learning Training (DLT) demands significant computational resources, leading to substantial energy consumption along with the associated financial costs and environmental effects. The computational demand for DLT continues to grow exponentially, driven by both the growing number of DL jobs-estimated to increase tenfold annually in Microsoft's production clusters \cite{gu2019tiresias}-and the need to scale computation for higher accuracy and capability. This surge is evident from the evolution of notable AI models like AlexNet in 2012 to AlphaGo Zero in 2019, representing a remarkable 300,000-fold increase in computational requirements \cite{openai_ai_compute}. 

In response, and to highlight this growing trend, several studies have measured and reported the energy consumption associated with various DLT tasks~\cite{patterson2021carbon, strubell2020energy, haghshenas2022co}. For instance, the energy usage of computation alone for training a Transformer-NAS on 8xV100 GPUs was reported to be 157.629 MWh~\cite{haghshenas2022co}. Such figures underscore the need for sustainable practices in deep learning, which can be addressed at multiple levels, from the design of Deep Neural Networks (DNNs) to system-level optimizations.

Running DLT on shared multi-tenant GPU clusters in private or public cloud environments, such as Google Cloud Platform (GCP)\cite{googlecloudgpu} and Microsoft Azure~\cite{microsoftdatabricks}, has become a standard practice in modern computing. Resource scheduling in such environments typically relies on traditional frameworks like Kubernetes~\cite{burns2016borg} and YARN~\cite{vavilapalli2013apache}, which were originally designed for generic workloads. These schedulers allocate GPU resources exclusively to each individual DLT job at its initiation and maintain exclusive access until job completion, which leaves resources underutilized or idle during periods of lower demand within jobs, such as during input preprocessing or I/O operations. 

Studies of production GPU clusters reveal significant under-utilization. For instance, Microsoft reports an average GPU utilization of only 52\% \cite{jeon2019analysis}, while Alibaba shows an even lower median utilization of just 10\% \cite{xiao2020antman}. Furthermore, an analysis of a two-month workload trace from another Alibaba GPU cluster indicates that the median usage of Streaming Multiprocessors (SMs) per task is around 0.042 GPUs \cite{weng2022mlaas}. Such under-utilization leads to inefficient resource usage, prolonged job waiting times, unfair job allocation, and increased energy consumption. Devices operating below their optimal capacity consume more energy relative to their work.

At the system level, addressing under-utilization is achievable by resource sharing, commonly referred to as consolidation, where resources are shared among multiple applications. With resource sharing, utilization, and performance change in a non-trivial manner~\cite{srikantaiah2008energy, haghshenas2019magnetic}. Consolidation leads to performance degradation due to contention among co-located applications for shared resources. Even in the absence of direct contention, resource sharing imposes coordination overhead, which can degrade performance. The objective is to balance the trade-off between optimizing resource utilization and maintaining acceptable performance levels. A fundamental performance metric for cloud platforms is meeting their Service Level Objective (SLO), specific numerical goals that define the availability, reliability, or performance of their systems. Any modification to system design must uphold these SLOs~\cite{googleaisla}. 

Although there has been research in GPU cluster management for DLT workloads, most relevant studies have thus far
been ad hoc, not guided by a rigorous analytical methodology. These efforts have primarily focused on optimizing
specific objectives, such as performance \cite{xiao2018gandiva, gu2019tiresias}, quality \cite{zhang2017slaq}, and fairness \cite{mahajan2020themis, chaudhary2020balancing}, without taking energy into account, and in some cases, resulting in increased energy consumption. Although some of these approaches incorporate GPU sharing as a mechanism \cite{xiao2018gandiva, wu2023transparent}, they typically employ it as a means to achieve their objectives without explicitly addressing energy efficiency.

In this paper, we investigate the dynamics of GPU sharing for DLT jobs and present the Energy-aware Co-Allocation (EaCO) algorithm, a new scheduling approach specifically designed to enhance energy efficiency in GPU clusters. By prioritizing energy-efficient scheduling, EaCO fills a critical gap in the current landscape of DLT workload management, making it the first framework to target this objective in GPU clusters. Our investigation focuses on key, unexplored questions in the literature: How does the co-location of DLT jobs influence energy efficiency, and under what conditions is it beneficial? Additionally, how does job co-location affect performance, and what trade-offs exist between energy efficiency and performance in this context? By answering such questions, we aim to provide a deeper understanding of the trade-offs and opportunities inherent in GPU sharing for DLT jobs.

To implement effective GPU consolidation, careful orchestration is required to minimize interference and contention for resources like memory, GPU compute cores, and interconnect bandwidth. DLT jobs from different applications, like many other cluster workloads, are heterogeneous in several aspects, including their resource requirements. This heterogeneity introduces specific challenges in GPU sharing as well as designing a scheduling algorithm that balances objectives while upholding job-specific constraints. To address this, workload profiling plays a crucial role to capture detailed information about job characteristics and resource usage patterns, enabling the scheduler to make informed decisions and to allocate resources dynamically based on job requirements and cluster conditions.

As it is impractical to profile all DLT jobs, EaCO uses a hybrid approach that combines profiling with history-based insights. Profiling provides fine-grained insights for known workloads, while history-based predictions leverage data from previously executed jobs to estimate the behavior of unseen workloads. Additionally, given that DLTs are inherently iterative, consisting of repeated stages of computation and communication, EaCO utilizes real-time metrics captured during the initial steps of job execution using monitoring tools. These early-stage observations reflect the job's steady-state behavior. By integrating this live feedback with profiling data and history-based predictions, EaCO adapts to the variability and complexity of DLT jobs, dynamically refining its scheduling decisions, and adapting to deviations from predicted behavior.

We implemented a profiling stage on servers equipped with NVIDIA V100 GPUs, focusing on image classification models. While this paper emphasizes these models, the methodology can be extended to other application domains. In addition, we integrated and evaluated EaCO within Gavel \cite{narayanan2020heterogeneity, GavelGitcode}, a cluster scheduler simulator designed for DLT workloads. The evaluation was conducted on clusters of two different scales: 28 nodes and 64 nodes, each with 8xV100 GPUs per node.
The results of our experiments demonstrate that co-locating jobs can improve energy efficiency by up to 44\% for single jobs (details in Sections 3 and 6.1).  
Simulations further show that EaCO reduces energy consumption by up to 39\% compared to existing scheduling algorithms when applied to job traces. If we define the Job Total Time (JTT) as the sum of job waiting time and job runtime, EaCO improves average JTT by 4.9\% to 97\%, depending on resource availability and workload demand (details in Section 6.2).

In summary, the main contributions of our work are as
follows.
\begin{itemize}
\item We experimentally demonstrate that co-locating DLT jobs can significantly improve energy efficiency and resource utilization, with negligible performance degradation compared to the typically long runtime of these jobs.

\item We characterize the behavior of DLT workloads under co-located execution, including its impact on key metrics such as resource utilization, Job Completion Time (JCT), job waiting time, and energy consumption.

\item We propose and evaluate EaCO, a novel scheduling framework that incorporates profiling-based insights to optimize the co-location of DLT workloads. EaCO refines scheduling decisions dynamically by leveraging historical data, estimation techniques, and early-stage observations. It achieves improved cluster and energy efficiency while meeting SLOs.

\end{itemize}

\section{Background}
\label{sec:background}

\subsection{Deep Neural Network Training}
DNNs consist of simple yet non-linear modules designed to transform raw input data into output representations. These networks rely on adjustable parameters known as weights. Typical DNNs have hundreds of millions of weights and hundreds of millions of raw examples that it is required to be trained~\cite{lecun2015deep}. 
The training process of a DNN involves many iterations, each handling a few samples of training data. In each iteration, a forward and backward pass is performed, requiring billions of floating-point operations~\cite{xiao2018gandiva}. Consequently, training of DNN models requires significant computation, which results in substantial energy consumption.

DNNs are widely applied across various domains, prompting the development of diverse architectures tailored to specific tasks. As a result, the computational requirements and resource utilization for training different architectures vary significantly. Resource utilization during the training process of DNNs depends on many factors, including the model and training configurations (e.g., network architecture, the number of parameters (weights), and hyperparameter settings), hardware setup, data complexity, and the degree of data or model parallelism. Additionally, the nature of training is stochastic to some extent coming from several stochastic elements such as random initialization. Therefore, the resource utilization as well as the Time to Target Accuracy (TTA) for DNN training jobs is hard to predict.

Among the various types of architectures, Convolutional Neural Networks (CNNs) are particularly prominent for applications like object detection and image classification. This paper specifically focuses on CNNs, however, the underlying principles and methodologies in this study are generalizable and broadly applicable to other neural network types and application domains.

\subsection{Optimizing Resource Utilization in GPU Clusters}

The training of DNNs heavily depends on advanced and expensive hardware platforms, such as Graphics Processing Units (GPUs) and Tensor Processing Units (TPUs), with GPUs being the primary choice. GPUs, like NVIDIA V100, differ from CPUs mainly due to their focus on parallel processing. The V100 has 80 SMs, each with 640 CUDA cores, resulting in a total of 5120 CUDA cores. This high level of parallelism allows the V100 to perform many calculations simultaneously. Additionally, its larger memory capacity compared to CPUs provides substantial acceleration in DNN training.

Context switching is a fundamental mechanism employed to switch between multiple processes, enabling their concurrent execution and improving overall resource utilization. In CPUs, context switching is handled by the operating system, while in GPUs, it is implemented directly at the hardware level. This hardware-based context switching in GPUs allows for latency hiding in a process by rapid switching between threads, typically executing within one or two cycles. Consequently, GPUs can manage more threads than the available Single Instruction-Multiple Data (SIMD) units. This capability of GPUs can be used by not only the application developers but also by higher abstraction level designers to run multiple applications and utilize available SMs more efficiently. 

This parallel processing capability of GPUs is further enhanced by their ability to handle numerous threads simultaneously. Efficient management of these threads is key to maintaining high utilization, and context switching plays a crucial role in this process. In a GPU, thread instructions are executed sequentially. As a result, to hide latencies and keep hardware busy, the execution of other warps (a warp being a set of 32 threads in CUDA) is essential when one warp is paused or stalled. Therefore, increasing the number of active warps generally improves GPU utilization. However, beyond a certain point, increasing occupancy no longer leads to performance gains, as the hardware resources become fully saturated \cite{nvidiacudaoccupancy}.
\section{Motivation}
\label{sec:motivation}

One promising strategy to reduce the energy usage of shared GPU/CPU clusters running DLTs is to aggregate tasks onto fewer active nodes. This minimizes job waiting times, and when there is no waiting, enables the activation of sleep modes or turning off idle devices. In contrast, the current method of managing DLTs is to assign dedicated GPUs for the job's entire duration. We argue against this approach and propose the sharing of GPUs among multiple DLT jobs for multiple compelling reasons, as discussed below.
\\

\noindent \textbf{A) Resource sharing improves energy efficiency and reduces waiting times}

We examine the potential of GPU sharing between DLT jobs to improve energy efficiency based on experimental results. In these experiments, we utilize servers equipped with 8 NVIDIA V100 GPUs to compare two distinct allocation strategies: exclusive assignment, referred to as \textit{no Space Sharing}, and \textit{Space Sharing}. The experimental setup and results are presented in detail in Section~\ref{sec:results_physical_server}.

\begin{figure}[!t]
\begin{center}
  \includegraphics[width=1\columnwidth, trim={0.2cm 0.2cm 0cm 0cm}, clip]{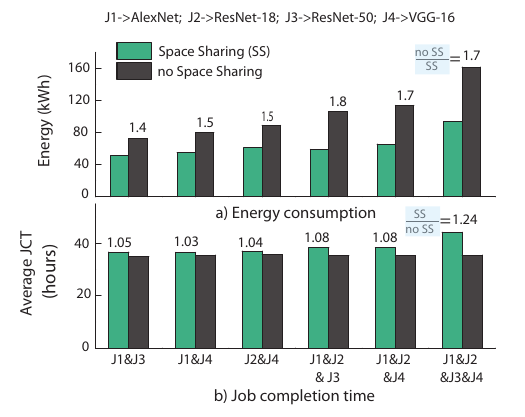}
\end{center}
\caption{\label{fig:energy&avgJCT} Total energy and average JCT for running a set of jobs, with and without space sharing.}
\end{figure}

Figure~\ref{fig:energy&avgJCT} illustrates the comparative analysis of total energy consumption and average JCT across various combinations of training jobs for multiple CNN models, including ResNet-18~\cite{he2016deep}, ResNet-50~\cite{he2016deep}, AlexNet~\cite{krizhevsky2012imagenet}, and VGG-16~\cite{simonyan2014very}.
Under the \textit{no Space Sharing} policy, each job is allocated exclusive access to a server node with 8xV100, keeping the resources until the job is completed. As an example, for the J2\&J4 job combination (where J2 corresponds to ResNet-18 and J4 to VGG-16, as indicated at the top of the figure), the \textit{no Space Sharing} policy dictates that J2 will run to completion on one server node, while J4 runs independently on another node. In contrast, the \textit{Space Sharing} policy, allows both J2 and J4 to execute simultaneously on the same server node.

As Figure~\ref{fig:energy&avgJCT}(a) shows, sharing resources improves energy efficiency by 30-44 percent across various job combinations. Under the \textit{Space Sharing} policy, co-locating jobs leverages the hardware execution of context switching which effectively mitigates latency during the transition between threads. Nonetheless, although minimized, this latency is not fully eliminated. Figure~\ref{fig:energy&avgJCT}(b) indicates that co-location increases average JCT by 3-19 percent compared to \textit{no Space Sharing} policy. 

Moreover, As more jobs are co-located, the system achieves higher energy efficiency by improving resource utilization, shown in Figure~\ref{fig:energy&avgJCT}. However, extensive co-location might result in substantial delays, imposing considerable performance degradation and, in some cases, increased energy consumption. In our experiments, the average JCT increase is kept below 8\% when two or three jobs are executed simultaneously, but it rises to 19\% when four jobs are co-located on a single node. Thus, GPU sharing requires careful consideration to balance energy efficiency and performance.  

In the real world, the assumption of having unlimited resources for exclusive allocation is rarely feasible due to the high demand for GPUs in shared clusters. Therefore, jobs often face significant queuing delays while waiting for dedicated resources to become available. Hence, enabling GPU sharing, not only enhances energy efficiency but also decreases job waiting periods and improves fairness within a shared cluster.\\

\noindent \textbf{B) Experiments highlight noticeable trends in resource utilization for DLT jobs}

As discussed, resource sharing offers significant advantages; however, exclusive allocation of DLT jobs remains a standard approach on many practical platforms. The main reason contributing to this preference is the difficulty in predicting TTA and resource utilization, specifically when co-locating jobs. Consequently, cluster orchestrator platforms prudently allocate these jobs to prevent failures originated from conflicts over resources \cite{zhang2020empirical}.

Recent studies on resource utilization prediction \cite{gao2020estimating, liu2022tbem} and runtime estimation \cite{bai2022dnnabacus, gao2023runtime} have introduced novel approaches, leveraging both analytical models and machine learning (ML) techniques. While these methods have shown high accuracy, generalizing such predictions to diverse environments remains a significant challenge due to several factors: (1) the training process of a neural network is stochastic to some extent, coming from several elements such as random initialization; (2) DNNs are highly heterogeneous in terms of architecture, application domains, and computational requirements; and (3) the introduced complexity of co-locating DLTs, where utilization and performance change in a nontrivial manner.

\begin{figure*}[!t]
\begin{center}
  \includegraphics[width=\textwidth, trim={0cm 2.7cm 0cm 0cm}, clip]{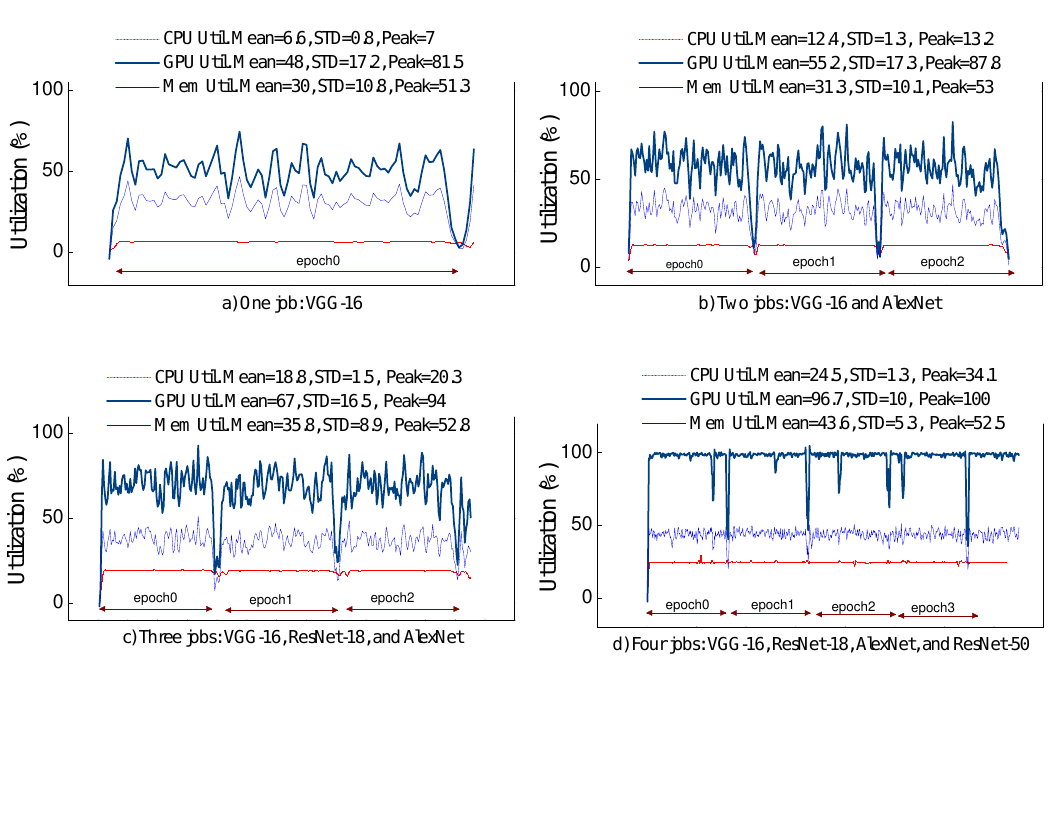}
\end{center}
\caption{\label{fig:utils} Resource utilization (GPU, CPU, and memory), while running different job combinations.}
\end{figure*}

While accurate prediction of resource utilization remains challenging; the results of our experiments reveal consistent trends in resource usage during the execution of both single and co-located DLT jobs. Figure~\ref{fig:utils} illustrates the utilization patterns of GPU, CPU, and memory across various combinations of DLT jobs during the initial epochs of the training process. The figure highlights an observable consistency in resource utilization trends across these epochs. Although fluctuations in utilization are evident within individual epochs, this behavior recurs with slight variations in subsequent epochs.

Additionally, a reasonable relationship appears to exist in the mean, standard deviation (STD), and, most importantly, peak utilization as more jobs are added. For example, mean CPU utilization increases progressively with each additional job, moving from 6.6\% to 12.4\%, then to 18.8\%, and finally reaching 24.5\%. GPU utilization shows a different pattern, rising from 48\% to 55.2\%, 67\%, and ultimately 96.7\%. These observations suggest that monitoring in the initial epochs would be beneficial for co-locating jobs, as DLTs are iterative and long-running, allowing early-stage trends to inform resource allocation strategies effectively. Finally, even in the experiment of co-locating four jobs, where peak GPU utilization reaches 100 percent, the execution of all four jobs continues reasonably without significant fluctuations.

What the results of this section suggest is that resource utilization while sharing resources is not completely unpredictable and exhibits distinguishable trends. Although predicting these patterns may pose challenges due to the inherent variability in job characteristics, the consistent relationships observed across different job combinations indicate that strategic monitoring and analysis of initial epochs can enhance resource allocation decisions. \\

\noindent \textbf{C) Available tools provide valuable real-time insights in GPU clusters}

Our experimental data provide valuable insights into the dynamics of resource utilization in shared GPU environments. A key question is how such insights can be effectively leveraged to optimize resource allocation strategies in GPU clusters. Given that DLTs are inherently repetitive and long-running, they are particularly suited for continuous monitoring and adaptive optimization. Real-time monitoring tools from NVIDIA, such as the Data Center GPU Manager~\cite{nvidiaDCGM}, System Management Interface (nvidia-smi)~\cite{nvidiasmi}, and Nsight Compute~\cite{nsightcompute}, play a crucial role in this process by providing fine-grained visibility into resource usage patterns.
 
DCGM is a tool developed by NVIDIA to monitor and manage GPU clusters in data centers. It provides detailed information on the utilization of GPU cores and memory within a cluster. One of its key functionalities is to assess computational occupancy~\cite{nvidiacudaoccupancy}, which refers to the amount of time a GPU is actively performing computations compared to its total available processing time. Metrics provided by monitoring tools along with predictions, allow administrators to optimize resource allocation, ensure efficient workload distribution, and maintain the high utilization of resources.

\section{System Model and Problem Formulation}
\label{sec:mechanisms}

In a shared GPU/CPU cluster, multiple server nodes, each equipped with several GPUs, are available to process incoming jobs. A scheduling algorithm assigns these GPUs to the tasks. The focus of this study is to minimize the energy consumption of executing a set of DLT jobs by optimizing computational resource allocation. Energy inefficiency in computation often arises from resource under-utilization, where nodes consume significant amounts of energy even when operating at low workloads or during idle periods. This is because the energy consumption of computational nodes does not scale linearly with workload reductions; a node operating below its capacity still consumes a large proportion of its peak energy. This imbalance leads to wasted energy if resources are not fully utilized. 

GPU sharing technique is a promising solution for this challenge to consolidate workloads on fewer active nodes. This consolidation allows waiting jobs to be scheduled more effectively and when no jobs remain in the queue, enables idle nodes to transition to low-power states. By leveraging these techniques, we aim to reduce overall energy consumption while maintaining the performance required for timely task completion.

This problem could be naively modeled as a multi-dimensional bin packing problem, treating servers as bins and each resource (e.g., GPU cores, memory, or bandwidth) as a separate dimension. However, while this abstraction captures the general resource allocation aspect, it fails to capture the unique challenges of GPU cluster environments such as performance degradation caused by co-allocation, power variation for different placements, and heterogeneity in hardware architectures across nodes, all of which influence energy usage and performance outcomes in complex ways \cite{srikantaiah2008energy}. Therefore, an effective optimization model must go beyond resource allocation to incorporate performance constraints and/or objectives. This ensures that energy savings are achieved without compromising the efficiency of job execution.

Co-allocating DLT jobs on shared GPUs often increases the average time per step/epoch compared to scenarios where jobs have exclusive access to the hardware. Additionally, maximizing GPU occupancy does not always correlate with improved throughput. Beyond a certain point, the increased load can result in diminishing returns or even negatively affect both energy efficiency and throughput due to factors like memory access delays, increased overhead for managing co-located workloads, and consequently reduced processing speed \cite{nvidiacudaoccupancy}.

In this work, we leverage job deadlines as representatives for SLOs, ensuring that all DLT jobs meet their performance expectations. we explicitly include deadline constraints in our problem formulation. Jobs are co-located only when the algorithm ensures that all deadlines are met. Incorporating performance as an explicit constraint within the optimization framework ensures that the scheduling algorithm is designed to prevent scenarios where co-location or resource contention compromises performance. Additionally, methods and techniques are integrated into the scheduling process to ensure that all SLOs are met, such as job deadlines or throughput requirements.

\subsection{Cost Function/Objective: Energy-Performance Trade-off} 
When sharing resources between multiple jobs, we encounter two scenarios. The first scenario occurs when there are ample resources available to exclusively allocate to all incoming jobs. In this case, co-location leads to a decrease in throughput, measured as the number of completed tasks over time. Thus, there exists a trade-off between energy consumption and throughput.

The second scenario arises when the available resources are insufficient to fulfill all jobs in the incoming queue. In this scenario, sharing resources improves throughput up to a certain point, by utilizing available resources more effectively, enabling more jobs to progress concurrently. In both scenarios, there is an inherent trade-off between energy usage and average time per step or epoch.

The objective of this work is to develop an energy-aware scheduling algorithm that optimally balances this trade-off. The cost function explicitly integrates energy consumption and average time per epoch (as the performance metric), as competing objectives. By dynamically adapting resource allocation strategies based on workload characteristics and cluster constraints, the algorithm aims to minimize energy usage while ensuring that performance remains within acceptable bounds, as defined by SLOs. We formulate the optimization problem as follows:

\begin{equation}
    \text{Minimize  } \alpha.\sum_{j=1}^{k} E_{j} + (1-\alpha). AvgTPE
\end{equation} 

\noindent where $E_{j}$ represents total energy for $j^{th}$ job, and $k$ denotes the number of all allocated jobs in $S_k$, the set of all allocated jobs. The second term, AvgTPE, represents the estimated average time per epoch (excluding waiting times) across all jobs in $S_k$. The parameter $\alpha$ determines the relative importance of energy efficiency and performance.

\subsection{Constraints: Deadlines and Overload}

In the design of load management systems, the SLOs play a crucial role in ensuring that task priorities and deadlines are met efficiently. SLOs define the expected performance criteria for a system, guiding how tasks are prioritized and allocated resources. Different systems, depending on their objectives and workload characteristics, establish specific commitments to meet the varying requirements of different job types and priorities. For instance, some tasks may require immediate completion, while others might have more flexible deadlines, for example within 12 hours. These less time-sensitive tasks can be treated as lower-priority jobs, and their allocation can be adjusted based on the availability of resources and the scheduling of higher-priority tasks.

In addition, priority calculation is a multi-faceted process, with each system using distinct criteria to rank tasks. For example, the Slurm cluster manager~\cite{yoo2003slurm} calculates job priority based on a weighted formula, factoring in elements like PriorityWeightAge (wait time), PriorityWeightFairshare (user resource balance), PriorityWeightQOS (Quality of Service), and PriorityWeightPartition (node group)~\cite{MultifactorPriorityPlugin}. Each factor's weight is configured by administrators to control scheduling behavior. The summed priority score determines the job order in the queue.

In this paper, we formalize the SLO by assigning specific deadlines to each job. Some jobs may have no explicit SLO, which we denote as a deadline of $\infty$, whereas high-priority jobs require immediate, exclusive resource allocation. To prevent deadline violations, we incorporate a constraint to the optimization problem. Specifically, if a scheduling decision results in an increase in JCT such that it exceeds the assigned deadline, the decision is discarded. We define the deadline constraint as follows:

\begin{equation}
      \text{s.t.  }\forall j \in S_k: \text{$deadline_j$ must be met.}\\
\end{equation} 

The next critical factor is to avoid overloading resources, mainly cores and memory. Therefore, the second and third constraints in the optimization problem are defined as follows:

\begin{equation}
      \text{s.t.  }\forall i \in S_g: \text{$U_i$} < U_{threshold}\\
\end{equation} 
\begin{equation}
      \text{s.t.  }\forall i \in S_g: \text{$U_{memi}$} < mem_{threshold}\\
\end{equation}

\noindent where $U_i$ and $U_{memi}$ represents GPU core and memory utilization of $i^{th}$ GPU in the set of all GPUs, $S_g$.

In the context of CPUs, achieving 100\% core utilization indicates that the entire capacity of the CPU cores is being utilized. Consequently, in CPU-only servers, a threshold is established for CPU utilization to prevent over-utilization. However, the same principle does not apply to GPUs. A GPU operating at 100\% core utilization does not necessarily imply full occupancy.

GPU utilization provided by monitoring tools like nvidia-smi \cite{nvidiasmi} is calculated by dividing the time which the GPU is active, running at least one kernel, by the total elapsed time since the last measurement. With this definition, GPU utilization is represented as 100\% if any SM remains active throughout the measurement interval, regardless of the number of SMs in use. This approach lacks granularity concerning how extensively the GPU's resources are employed, as high utilization values may be recorded even when most SMs are idle.

In contrast, occupancy provides a finer-grained metric, which measures the proportion of active warps per multiprocessor in relation to the hardware’s maximum warp capacity. High occupancy is desirable as it indicates effective use of the GPU’s parallel processing capabilities, thus optimizing hardware resources through efficient employment of threads and blocks.
Hardware occupancy can be improved by concurrent kernel execution which allows multiple independent kernels to execute concurrently \cite{nvidiacudabestpractices}. When an application does have enough concurrent kernels to use all the GPU resources, the resource occupancy can be further increased by assigning multiple applications to one GPU, ensuring to have enough kernels. 

In this work, we opt to use GPU utilization as a conservative measure, while occupancy may offer more energy-efficient solutions through increased co-location of tasks. By using utilization as a metric, we maintain a balance between effective resource usage and performance, preventing potential performance degradation while still optimizing energy consumption.
\section{EaCO: Proposed Energy-aware CO-allocating Algorithm}
\label{sec:proposedapproach}

In dynamic environments such as cloud clusters, load scheduling is a complex optimization problem under uncertainty. This environment is characterized by features like unpredictable job arrivals, fluctuating computational demands, and limited foresight regarding resource availability, all of which make traditional optimization approaches computationally expensive and often impractical. Therefore, greedy and suboptimal algorithms are commonly used to schedule and allocate jobs. Greedy algorithms provide a pragmatic solution by making locally optimal choices at each decision step, thereby enabling quick, real-time responses to changes in workload patterns and resource demands. This approach prioritizes quick solutions over achieving globally optimal ones, acknowledging the challenges posed by uncertain and evolving conditions within GPU cluster environments.

This section introduces our proposed scheduling solution, EaCO, which addresses the key question: \textit{How can we co-allocate DLT jobs to improve energy efficiency while maintaining performance guarantees?}
EaCO is designed based on experimental evaluations of co-locating different job combinations. Given the variety of job types, possible configurations, and possible combinations, conducting a complete set of experiments is not practical. Therefore, EaCO incorporates a predictive approach to estimate the performance of unseen job sets, utilizing both experimental data and historical records that accumulate over time. As these predictions may be somewhat inaccurate, EaCO also incorporates early-stage observations. As shown in Figure \ref{fig:utils}, the performance observed in the initial epochs tends to repeat in subsequent epochs. Therefore, using early-stage observations, EaCO can achieve more precise estimates of execution times, allowing it to evaluate and refine job allocations to ensure optimized energy-performance trade-off across co-allocated tasks.

Algorithm~\ref{alg1} reviews the main components of the EaCO algorithm as pseudo-code, which we assume to have deadlines known for incoming jobs. The algorithm operates by dynamically adjusting allocation decisions based on real-time data about job characteristics and resource availability.

EaCO begins the allocation process for job $j$ based on a scheduling order determined by the arrival time and priority of jobs in the incoming queue. The algorithm leverages a history of experimental measurements to improve its prediction accuracy. A larger data history allows it to make faster and more accurate estimates. The first step in allocating job $j$ includes identifying candidate GPU sets that meet $j$'s memory and core requirements, while also considering real-time GPU utilization, a process handled by the \textbf{FindCandidates} function (Line 2).

Once the list of potential candidates $L$ is given, EaCO tries to finalize the allocation of $j$ through two nested while loops (Lines 3–21). Within these loops, the algorithm selects the GPU set $G$ with the highest utilization from $L$ and employs the \textbf{PredictJCT} function (Line 6) to estimate the JCT for all jobs co-located on $G$, including $j$. If this estimation indicates that the deadlines for all jobs, including $j$ will be met under the proposed allocation, then EaCO proceeds with the allocation of $j$ to $G$.  

To finalize the allocation of $j$, EaCO employs early-stage observations by continuously monitoring the performance of all co-located jobs on the selected GPU set $G$.
As each job completes one epoch, EaCO can refine its JCT estimation to achieve a higher accuracy in its predictions.
If the updated estimation indicates that a deadline for any co-located job will not be met, the allocation of $j$ to $G$ is immediately reversed to avoid potential deadline violations. This monitoring process continues until one epoch has passed for every job co-located on $G$, allowing EaCO to make a reliable decision. If the deadlines for all jobs continue to be achievable under the current allocation, EaCO then finalizes the allocation of $j$ to $G$, ensuring efficient resource utilization without compromising on job deadlines. 

EaCO ensures that allocation reversals occur only at the end of epochs. This guarantees that any interruption in the training process happens at a natural checkpoint, where the model's progress is saved. Consequently, if an allocation is undone, the training resumes from the last saved point, minimizing overhead to JCT and ensuring that no computational effort is wasted.

\begin{algorithm}[!t]
\floatname{algorithm}{\textbf{Algorithm}}
\small
\caption{\small EaCO: Energy-aware CO-allocating algorithm}
\label{alg1}
\renewcommand{\algorithmicrequire}{\textbf{Schedule} (job $j$)}

\begin{algorithmic}[1]
\REQUIRE 
\STATE Initialize history $H$ with experimental measurements.
\STATE $L$ $\leftarrow$ list of candidate GPU sets based on job requirements and observed utilization (below a predefined threshold) by \textbf{FindCandidates} (see Algorithm~\ref{alg2}).

\WHILE{allocation for $j$ is not finalized}
\WHILE{$j$ is not allocated}
\STATE $G \leftarrow$ GPU set in $L$ with highest utilization 
\STATE Predict JCT for all co-located jobs on $G$ and $j$ using $H$.
\STATE Remove $G$ from $L$.
\IF{deadline is met for all jobs}
    \STATE Allocate $j$ to $G$.
\ENDIF
\ENDWHILE

\STATE Continuously profile performance for all co-allocated jobs on $G$ until one epoch has passed for all co-located jobs since allocation.
\STATE Record measured performance in history $H$.
\STATE Estimate JCT for all co-located jobs.
\IF{deadline is met for all jobs} 
    \STATE Finalize allocation.
\ELSE
    \STATE Undo the allocation.
    \STATE Go to Line 3.
\ENDIF
\ENDWHILE
\end{algorithmic}
\end{algorithm}

\textbf{Resource provisioning: }EaCO uses the FindCandidates function (described in Algorithm~\ref{alg2}) to select potential GPU sets for job placement, considering both the job’s requirements and the availability of resources. For each job, GPU type as well as memory and core utilization are the primary factors to match the resources. When placing a single-GPU job, the primary factors are the current utilization and type of the GPU, however, in multi-GPU jobs, additional considerations emerge, such as whether the GPUs are on the same physical node. While these aspects are critical, they fall outside the scope of EaCO's focus, which is optimizing GPU sharing.

EaCO’s FindCandidates function ensures that estimated memory is fully allocated, only permitting GPU sharing when the available memory exceeds the cumulative demand from all co-located jobs. Previously developed models like DNNMem~\cite{gao2020estimating} and TBEM~\cite{liu2022tbem} have shown effectiveness in estimating the memory usage of DLTs. In co-location assignments, we consider the GPU's peak memory usage for currently assigned jobs and apply a predefined threshold to minimize the probability of memory contention. 

EaCO sets core and memory utilization thresholds and filters potential GPUs for co-allocation; GPUs operating below these thresholds are eligible candidates, while those above are excluded. Algorithm~\ref{alg2} reviews the procedure of the FindCandidates function.

\begin{algorithm}[!t]
\floatname{algorithm}{\textbf{Algorithm}}
\small
\caption{\small FindCandidates: Finding potential candidates}
\label{alg2}
\renewcommand{\algorithmicrequire}{\textbf{FindCandidates }(job $j$)}
\begin{algorithmic}[1]
\REQUIRE 
\STATE $All\_subsets$ $\leftarrow$ All subsets of GPUs based on ($j$'s requested GPUs.
\FOR{$S$ in $All\_subsets$ }
\FOR{$g$ in $S$ }
\IF{($U_g$>$U_{threshold}$) \OR ($U_{memg}$>$mem_{threshold}$) }
    \STATE break
\ENDIF
\STATE$avail\_mem$ += $g.total\_mem$ - $g.peak\_usage$
\ENDFOR
\IF{$avail\_mem$ > $j$'s estimated memory}
    \STATE Add $S$ to $candidates$.
\ENDIF
\ENDFOR
\RETURN $candidates$

\end{algorithmic}
\end{algorithm}

In the FindCandidates function, a job $j$ is received as input, and the function iterates over all possible GPU sets for allocation (Lines 2-12). These sets are determined based on the number and type of GPUs requested by $j$ (Line 1).

For each potential set, while iterating over its GPUs (Lines~3-8), the core and memory utilization of each GPU are evaluated against a predefined threshold. If any GPU within the set exceeds this utilization threshold, the loop is interrupted, and the function proceeds to the next possible set. If not, the available memory across all GPUs in the set is accumulated. As depicted in Figure \ref{fig:utils}, memory utilization fluctuates during model training. Consequently, available memory is computed based on the peak memory usage of the jobs currently allocated to the GPU.

If the total available memory across the GPUs meets the memory required for job $j$, the GPU set $S$ is added to the list of candidates. For multi-GPU jobs, resource management can be handled differently, depending on the training approach and data distribution technique, such as model parallelism and data parallelism. This also affects memory usage across GPUs. Here, Algorithm~\ref{alg2} assumes model parallelism, where different GPUs are responsible for storing and processing distinct portions of the DNN.
\section{Evaluation}
\label{sec:evaluation}

In this section, we first experimentally examine the energy usage of computation for training various DNN models and their various combinations when co-located. Then we evaluate our proposed approach for large-scale clusters by simulations.

\subsection{Results from Cluster Experiments}
\label{sec:results_physical_server}

For our experiments, we used servers with 2x Intel Xeon Gold 6240 (36-cores total @2.60GHz) accompanied by 768GB RAM and eight V100 GPUs. We use four known Computer Vision (CV) models, AlexNet, ResNet-18, ResNet-50, and VGG-16 trained with ImageNet dataset \cite{deng2009imagenet}. The models are implemented in PyTorch and their code is freely available and we have used them out-of-the-box \cite{pytorch_imagenet_examples}. All the models have been trained in the experiments with batch size and the number of epochs/steps reported in their original papers. We train the models described with the default configuration and measure the power and JCT of each job.
 
The power consumption of a DLT job is determined by aggregating the power contributions of its hardware components, including GPUs and server infrastructure. For GPU power measurement, we use the $nvidia-smi$ interface, which provides real-time power draw metrics for each GPU. These are sampled at regular intervals to capture the dynamic power consumption during the training process. For the server's power consumption, we adopt a power model approach by estimating it based on CPU utilization~\cite{fan2007power}. Finally, the power of a DLT job at time $t$ is calculated as follows~\cite{haghshenas2022co}:

\begin{equation}
\label{job_power}
    P_{(j, t)} = P_{(server, t)} + \sum_{g=1}^{g=n} P_{(g, t)}
\end{equation} 
where $P_{(g, t)}$ and $n$ stand for the power of $g^{th}$ GPU and the number of GPUs used for the training, respectively.

The epoch time, JCT, average power, and total energy for referenced models are presented in Table~\ref{tab:power}. In addition, Table~\ref{tab:utils} summarizes the average and maximum memory and GPU utilization across all GPUs employed during the model training processes.

CPU and GPU utilization are the most indicative factors influencing a job's power consumption. The measurements show significant variability in resource utilization, as well as the corresponding power demands for training different models. This highlights the energy and resource usage inefficiency associated with exclusive allocation strategies of DLT jobs. It also underscores opportunities for optimizing resource usage by co-allocating various jobs with distinct resource requirements and usage patterns. 

\begin{table}[!b]
\caption{Average power, total energy, epoch time, and JCT for training different models.} 
\captionsetup{position=top}
\centering 
\label{tab:power}
\renewcommand{\arraystretch}{1.2}
  \begin{tabular}{|l c c c c|}
    \hline
     \footnotesize\textbf{Job} &  \footnotesize\textbf{\begin{tabular}[c]{@{}c@{}}{Avg.}\\{Power (W)}\end{tabular}} & \footnotesize\textbf{\begin{tabular}[c]{@{}c@{}}{Tot.Energy}\\{(kWh)}\end{tabular}} &
     \footnotesize\textbf{\begin{tabular}[c]{@{}c@{}}{JCT}\\{(H)}\end{tabular}} & \footnotesize\textbf{\begin{tabular}[c]{@{}c@{}}{Avg.Epoch}\\{Time (H)}\end{tabular}} \\ \hline \noalign{\hrule height 1pt}

     {\footnotesize{AlexNet}} & \footnotesize{712} & \footnotesize{24.73}&  \footnotesize{34.76} & \footnotesize{0.39}\\ \hline
     {\footnotesize{ResNet-18}} & \footnotesize{959} & \footnotesize{33.69} & \footnotesize{35.13}&  \footnotesize{0.39}\\ \hline
     {\footnotesize{ResNet-50}} & \footnotesize{1330} & \footnotesize{47.87} & \footnotesize{36.01} & \footnotesize{0.4}\\ \hline
     {\footnotesize{VGG-16}} & \footnotesize{1533} & \footnotesize{55.38}& \footnotesize{36.13} & \footnotesize{0.4}\\ \hline

  \end{tabular}
\end{table}

\begin{table}
\caption{Average and maximum GPU resource utilization during model training.} 
\centering 
\label{tab:utils}
\renewcommand{\arraystretch}{1.2}
  \begin{tabular}{|l c c c c |}
    \hline
     \footnotesize\textbf{Job} & \footnotesize\textbf{\begin{tabular}[c]{@{}c@{}}{Avg. Mem}\\{Util (\%)}\end{tabular}} & \footnotesize\textbf{\begin{tabular}[c]{@{}c@{}}{Max. Mem}\\{Util (\%)}\end{tabular}} & \footnotesize\textbf{\begin{tabular}[c]{@{}c@{}}{Avg. GPU}\\{Util (\%)}\end{tabular}} & \footnotesize\textbf{\begin{tabular}[c]{@{}c@{}}{Max. GPU}\\{ Util (\%)}\end{tabular}} \\ \hline \noalign{\hrule height 1pt}

     {\footnotesize{AlexNet}} & \footnotesize{1.73} & \footnotesize{4.21} &\footnotesize{4.72} & \footnotesize{11}\\ \hline
     {\footnotesize{ResNet-18}} & \footnotesize{6.07} & \footnotesize{14.63} &\footnotesize{11.17} & \footnotesize{27.29} \\ \hline {\footnotesize{ResNet-50}} & \footnotesize{22.29} & \footnotesize{43.92} & \footnotesize{36.61} & \footnotesize{72.04} \\ \hline
     {\footnotesize{VGG-16}} & \footnotesize{30.03} & \footnotesize{51.29} &\footnotesize{48.01} & \footnotesize{81.5}\\ \hline

  \end{tabular}
\end{table}

In subsequent experiments, we evaluate the impact of GPU sharing. A part of the results, presented in Figure~\ref{fig:energy&avgJCT}, demonstrate that resource sharing reduces total energy by 30-44\% compared to exclusive allocation. Table \ref{tab:power_co_allocated} represents the average power, epoch time, and JCT, as well as total energy, for running various job sets. Despite the efficiency of context switching in GPUs, switching latency is not zero, and even with having enough resources available, when multiple DLT jobs run on the same set of GPUs, it imposes some delay overheads. This delay leads to increased epoch time and consequently JCT.

Sharing GPUs among two or three concurrent jobs resulted in a performance decrease of 3 to 7.8 percent compared to running each job exclusively on a dedicated GPU set. In these tests, the program interchanges between jobs at each training step, sequentially executing them. However, with four co-allocated jobs, the program behavior changes, no longer following a strictly sequential switching pattern. Consequently, certain jobs occupy the GPUs for extended durations, making it unfeasible to record precise epoch times for each individual job. Nevertheless, when evaluating the overall effect across all co-allocated jobs, JCT increased by 19\% relative to exclusive access. 

As depicted in Figure~\ref{fig:energy&avgJCT}, running four jobs concurrently reduces total energy consumption by 42 percent. Whether this trade-off is acceptable or not depends on the performance requirements and the SLO specifications, which may prioritize energy efficiency or specific latency thresholds.

\begin{table}
\caption{Average power, total energy, average epoch time, and average JCT across co-allocated jobs.} 
\centering 
\label{tab:power_co_allocated}
\renewcommand{\arraystretch}{1.2}
  \begin{tabular}{|l c c c c|}
    \hline
     \footnotesize\textbf{Job Set} & 
     \footnotesize\textbf{\begin{tabular}[c]{@{}c@{}}{Avg.}\\{Power (W)}\end{tabular}}& 
     \footnotesize\textbf{\begin{tabular}[c]{@{}c@{}}{Tot.Energy}\\{(kWh)}\end{tabular}} &
     \footnotesize\textbf{\begin{tabular}[c]{@{}c@{}}{Avg.}\\{JCT (H)}\end{tabular}}& 
     \footnotesize\textbf{\begin{tabular}[c]{@{}c@{}}{Avg. Epoch}\\{Time (H)}\end{tabular}} \\ \hline \noalign{\hrule height 1pt}

     {\footnotesize{J1\&J3}} & \footnotesize{1390}& \footnotesize{50.93}& \footnotesize{36.63} & \footnotesize{0.407} \\ \hline
     {\footnotesize{J1\&J4}} & \footnotesize{1506}& \footnotesize{54.97}& \footnotesize{36.51} & \footnotesize{0.406}\\ \hline
     {\footnotesize{J2\&J4}} & \footnotesize{1644}& \footnotesize{60.84}& \footnotesize{37.01} & \footnotesize{0.411} \\ \hline
     {\footnotesize{J1\&J2\&J3}} & \footnotesize{1541}& \footnotesize{59.01}& \footnotesize{38.28} & \footnotesize{0.425} \\ \hline
     {\footnotesize{J1\&J2\&J4}} & \footnotesize{1713}& \footnotesize{65.55}& \footnotesize{38.26} & \footnotesize{0.425} \\ \hline  
     {\footnotesize{J1\&J2\&J3\&J4}} & \footnotesize{1944}& \footnotesize{93.66}& \footnotesize{44.21} & \footnotesize{-} \\ \hline
    \multicolumn{5}{c}{J1->AlexNet, J2->ResNet-18, J3->ResNet-50, J4->VGG-16}\\
  \end{tabular}
\end{table}

Finally, Table~\ref{tab:utils_co_allocated} presents the mean and maximum utilization of memory and core resources across all employed GPUs during various job combinations. When compared to the data in Table \ref{tab:utils}- which details resource usage under exclusive allocation— both the mean and maximum memory utilization in the co-allocated scenario are lower than the aggregated values of those in the exclusive scenarios. However, for core utilization, this trend is not consistent; while generally lower, in some cases, the average in the co-allocated scenario is higher than the aggregated (up to 4.6 percent in our experiments). This observation underscores the need for early-stage feedback, especially regarding core utilization during co-allocation. Notably, maximum GPU utilization reached 100\% only with four concurrent jobs, yet no job failures occurred, and all co-allocated tasks completed successfully.

\begin{table}
\caption{Average and maximum GPU resource utilization while running co-allocated jobs.} 
\centering 
\label{tab:utils_co_allocated}
\renewcommand{\arraystretch}{1.2}
  \begin{tabular}{|l c c c c |}
    \hline
     \footnotesize\textbf{Job Set} & 
     \footnotesize\textbf{\begin{tabular}[c]{@{}c@{}}{Avg. Mem}\\{Util (\%)}\end{tabular}} & 
     \footnotesize\textbf{\begin{tabular}[c]{@{}c@{}}{Max. Mem}\\{Util (\%)}\end{tabular}} & 
     \footnotesize\textbf{\begin{tabular}[c]{@{}c@{}}{Avg. GPU}\\{Util (\%)}\end{tabular}} & 
     \footnotesize\textbf{\begin{tabular}[c]{@{}c@{}}{Max. GPU}\\{ Util (\%)}\end{tabular}} \\ \hline \noalign{\hrule height 1pt}

     {\footnotesize{J1\&J3}} & \footnotesize{22.66} & \footnotesize{46.25} &\footnotesize{40.25} & \footnotesize{76.67}\\ \hline
     {\footnotesize{J1\&J4}} & \footnotesize{31.26} & \footnotesize{52.96} &\footnotesize{55.16} & \footnotesize{87.75} \\ \hline 
     {\footnotesize{J2\&J4}} & \footnotesize{34.85} & \footnotesize{52.54} & \footnotesize{61.06} & \footnotesize{93.46} \\ \hline
     {\footnotesize{J1\&J2\&J3}} & \footnotesize{27.77} & \footnotesize{55.88} &\footnotesize{52.24} & \footnotesize{91.88}\\ \hline
     {\footnotesize{J1\&J2\&J4}} & \footnotesize{35.83} & \footnotesize{52.75} &\footnotesize{66.99} & \footnotesize{93.96}\\ \hline 
     {\footnotesize{J1\&J2\&J3\&J4}} & \footnotesize{43.46} & \footnotesize{52.54} &\footnotesize{96.64} & \footnotesize{100	}\\ \hline
     \multicolumn{5}{c}{J1->AlexNet, J2->ResNet-18, J3->ResNet-50, J4->VGG-16}\\
  \end{tabular}
\end{table}

\subsection{Results from Simulations}

To evaluate the potential energy efficiency improvement achievable by GPU sharing within large-scale GPU clusters, we employed Gavel simulator. The simulator, in its original configuration, does not have the capabilities for these evaluations. Consequently, we (1) integrated the empirical data and measurements derived from our experiments on power consumption and performance into the simulator, and (2) developed essential modules to enable the computation of energy usage and other relevant metrics, as outlined in the subsequent sections.

We compare our proposed scheduling algorithm, EaCO, against three existing algorithms in the Gavel simulator: default, FIFO\_packed, and Gandiva. The default is a basic First-In, First-Out (FIFO) scheduler that assigns jobs sequentially and exclusively based on their arrival order. FIFO\_packed also follows a FIFO approach but allows packing jobs if there is not enough space for new jobs. Gandiva is an advanced scheduling algorithm from a state-of-the-art study \cite{xiao2018gandiva} that incorporates job packing within a greedy scheduling framework to optimize resource utilization.

\begin{figure}[!t]
\begin{center}
  \includegraphics[width=1\columnwidth, trim={0cm 0cm 0cm 0cm}, clip]{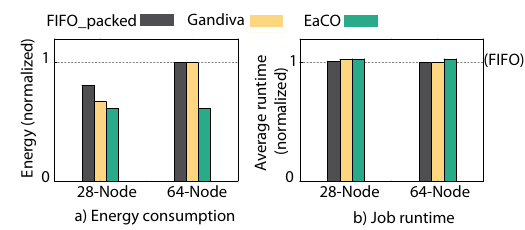}
\end{center}
\caption{\label{fig:energy&avgJCT&gavel} Total energy and average job runtime for executing a job trace, normalized to the default FIFO algorithm, across two cluster sizes.}
\end{figure}

Figure~\ref{fig:energy&avgJCT&gavel}(a) represents the total energy consumption of a GPU/CPU cluster, when running a workload trace with various scheduling algorithms, normalized to the default FIFO algorithm. The results include two cluster configurations: a 28-node cluster and a 64-node cluster, with each server node equipped with 8 GPUs. When the capacity of the cluster is higher than the workload demand (as in the 64-node configuration in our simulations), both FIFO\_packed and Gandiva exhibit comparable energy consumption, as no job packing is necessary. In contrast, EaCO achieves a significant reduction in total energy, decreasing it by 39\% compared to the other three algorithms in the 64-node cluster configuration.

In scenarios with constrained cluster capacity, such as the 28-node cluster configuration in our simulations, FIFO\_packed and Gandiva start packing jobs, resulting in lower energy consumption compared to the default FIFO. Nevertheless, even in this scenario, EaCO achieves the lowest energy consumption. Specifically, EaCO reduces energy by 39\%, 24.5\%, and 8.3\% compared to the default, FIFO\_packed, and Gandiva algorithms, respectively.

As discussed earlier, while co-locating jobs improves energy efficiency, it causes a slight performance trade-off in the form of increased individual job runtimes. Figure~\ref{fig:energy&avgJCT&gavel}(b) illustrates the average job runtime with various algorithms, normalized to the default. EaCO increases the average job runtime by less than 3.23\% compared to other algorithms, a marginal overhead given the long runtimes of DLT jobs. For example, in our experiments, the baseline runtime of a job like ResNet-18 training on a node with 8 V100 GPUs is approximately 36.13 hours. With EacO, this may increase to a worst-case runtime of 37.21 hours, an increase of just over an hour. Given that these long-running training jobs are typically executed in batch mode and are not time-critical, this increase in runtime is considered negligible.

Additionally, our performance-aware packing algorithm significantly reduces job waiting times, particularly in scenarios where the system capacity does not meet the demand. By efficiently packing jobs, EaCO ensures that resources are utilized optimally, thus minimizing idle time. It decreases the average JTT by 97\% in the 28-node configuration when compared to the default scheduler.

\begin{figure}[!t]
\begin{center}
  \includegraphics[width=1\columnwidth, trim={0cm 0cm 0cm 0cm}, clip]{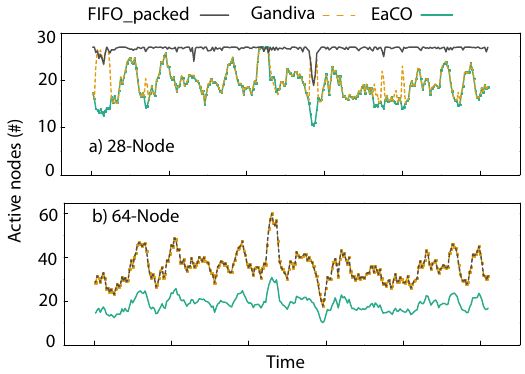}
\end{center}
\caption{\label{fig:active_nodes} Number of active nodes employing different algorithms for a) 28-Node b) 64-Node cluster configurations.}
\end{figure}

Finally, the cluster scheduler can significantly impact the number of active nodes required at any given time. As depicted in Figure~\ref{fig:active_nodes}, for both 28 and 64-node configurations the default scheduler uses the maximum number of nodes, which is expected since it does not optimize for resource utilization. 

FIFO\_packed also utilizes all available nodes for job allocation. However, its number of active nodes is slightly lower than that of the default scheduler when some jobs are completed, freeing up their respective nodes. As a result, during the gap between job arrivals, some nodes remain idle until new jobs are scheduled. The EaCO algorithm, on the other hand, demonstrates a marked improvement in resource utilization, reducing the average number of active nodes in a 28-node cluster by 30\%, 29\%, and 4\% compared to the default, FIFO\_packed, and Gandiva schedulers, respectively.

The reduction in the number of active nodes achieved by EaCO is particularly significant in terms of energy efficiency. Despite the decrease of only 4\% in the number of active nodes compared to the Gandiva scheduler, EaCO manages to reduce the total energy consumption by 8.3\%. This improvement in energy efficiency is attributed to EaCO’s ability to reduce waiting times and its strategy of co-allocating tasks more effectively, resulting in a substantial reduction in the overall runtime of the workload trace. Unused nodes can either be allocated additional load or placed in low-power states, contributing to the overall energy savings.

In the case of the 64-node cluster, where the number of available GPUs is much higher than the immediate load, all three algorithms—default, FIFO\_packed, and Gandiva—similarly assign jobs, resulting in an identical number of active nodes. This is expected, as the resource availability allows them to allocate jobs without optimizing for node utilization. However, the EaCO algorithm demonstrates a notable improvement, reducing the average number of active nodes by 47\%.
\section{State of the Art}
\label{sec:relatedwork}

In this section, we review several existing schedulers designed for DNN training workloads, each targeting one or more objectives including model quality, performance (e.g., makespan or JCT), and fairness.

\textbf{Quality-oriented:} 
SLAQ explores the quality-runtime trade-off across multiple DLT jobs, aiming to maximize system-wide model quality \cite{zhang2017slaq}. Quality, in this context, measures how accurately a model maps inputs to target outputs. SLAQ monitors the history of quality and allocates more resources to the jobs that have more potential for quality improvement. Similarly, OASiS 
schedules the arriving jobs while dynamically adjusting the numbers of concurrent workers and parameter servers during runtime \cite{bao2018online}. Like SLAQ, OASiS focuses on improving model quality without considering other performance metrics.

\textbf{Performance-oriented:}
Many schedulers prioritize optimizing performance metrics such as JCT. SLearn 
employs task sampling to estimate and optimize the average JCT for a trace of jobs. This approach leverages similarities in runtime properties among tasks within the same job \cite{jajoo2022slearn}. Similarly, Marcelo et al. developed a topology-aware scheduling algorithm that considers hardware communication requirements to improve execution time \cite{amaral2017topology}.

Tiresias schedules and places distributed DNN training jobs in order to minimize JCT, balancing user-centric and operator-centric goals (minimizing JCT and maximizing GPU utilization, respectively) \cite{gu2019tiresias}. Le et al. have proposed Allox that optimizes JCT while ensuring fairness among users by transforming the multi-configuration job scheduling problem into a min-cost bipartite matching solvable in polynomial time \cite{le2020allox}. These schedulers target performance optimization, often at the expense of resource and energy usage efficiency.

\textbf{Efficiency-oriented:}
Approaches focusing on efficiency leverage co-location to improve resource utilization. These methods operate at different abstraction levels including OS and application. 

Gandiva redefines the scheduling unit from entire jobs to micro-tasks, enabling the oversubscription of DLT jobs and providing early feedback \cite{xiao2018gandiva}. Similar to SLearn~\cite{jajoo2022slearn}, Gandiva exploits early feedback to perform profile-driven introspection and uses the mini-batch progress rate to make its allocation decisions, optimizing cluster efficiency. Operating at the application layer, Salus optimizes JCT and resource utilization using fast job switching and centralized memory management. These features allow it to pack more small DLT jobs on the same device \cite{yu2019salus}, thereby increasing system throughput. Antman 
is another solution that co-designs cluster scheduler and DL framework to enable cooperative job management to improve overall system efficiency.

PipeSwitch minimizes switching delays using pipelined context switching \cite{bai2020pipeswitch}. With PipeSwitch only one job can execute at any time. Orion operates at the level of individual DNN operators, reducing interference by considering the compute and memory requirements of each operator \cite{strati2024orion}. Finally, Liquid by Gu et al. introduces a regression-based method to estimate the resource requirements of DLTs, thereby avoiding resource over-allocation~\cite{gu2021liquid}. It integrates an intelligent, network-efficient cluster scheduler to minimize JCT. Liquid supports fine-grained GPU sharing, enhancing both resource utilization and system throughput.

All these approaches integrate co-location (also referred as "packing" and "GPU sharing"), as a potential scheduling mechanism. However, this action is utilized in scenarios where resource availability is not sufficient.

\textbf{Fairness-oriented:}
Gandivafair by Chaudhary et al. builds on Gandiva by balancing cluster efficiency and fairness across CPU/GPU resources \cite{chaudhary2020balancing}. It ensures a fair distribution of heterogeneous GPUs across users while redistributing unused GPUs to active users. THEMIS also targets fairness and implements a two-level algorithm where DLT workloads bid for resources based on recommendations from a central arbiter \cite{mahajan2020themis}. This algorithm allocates GPUs to winning bids, considering trade-offs between fairness and JCT. 

All the approaches mentioned above are energy oblivious and focus on performance-related objectives. However, the race for performance and accuracy in training DNN models comes with both financial and environmental costs, an aspect that has been overlooked in previous work. Meanwhile, Gandiva is the closest to the scope of our work, as it uses packing and early feedback strategies. Therefore, we include Gandiva in our evaluations to examine the effectiveness of EaCO.

\section{Conclusion}
\label{sec:conclusions}

This paper presents an experimental and simulation-based evaluation of the impact of GPU sharing on energy efficiency, resource utilization, and performance. We introduce EaCO, a new scheduling algorithm designed for the co-allocation of DLT jobs in shared GPU clusters, with the primary goal of improving resource and energy utilization while ensuring job deadlines. EaCO leverages historical performance data and predicts the performance of new job allocations, allowing for more informed scheduling decisions. Additionally, the algorithm uses early-stage observations to proactively avoid deadline misses. The results demonstrate that EaCO improves both energy and resource utilization compared to traditional scheduling methods, while effectively maintaining performance.

\bibliographystyle{acm}
\bibliography{biblio}

\end{document}